\documentclass[aps,onecolumn,showpacs]{revtex4}
\usepackage{amsmath,graphicx,subfig}
\usepackage{caption}
\usepackage{amssymb,amsbsy}

\begin{document}

\begin{center}{\large{\bf Duality invariance in massive theories }}
\vskip 0.5cm
Adel Khoudeir
\\ \vskip 0.2cm
{\it Centro de F\'{\i}sica Fundamental, Departamento de F\'{\i}sica, \\
Facultad de Ciencias, Universidad de Los Andes, M\'erida 5101, Venezuela}
\vskip 0.3cm
{\it e-mail: adel@ula.ve}
\begin{center}
{and}
\end{center}
\vskip 0.5cm
David Sierra
\\ \vskip 0.2cm
{\it Laboratorio de Astronom\'{\i}a y F\'{\i}sica Te\'{\o}rica, LAFT. \\ Facultad Experimental de Ciencias. Universidad del Zulia. Maracaibo 4001, Venezuela}.
\vskip 0.3cm
{\it e-mail: dsierra@fec.luz.edu.ve }
\end{center}

\vskip 1cm

\begin{center}
ABSTRACT
\end{center}
In this work, we show that duality symmetry can be implemented for massive theories at the level of the action, whenever we
can formulate appropriates gauge invariant actions. For a massive vectorial field, we use a known gauge invariant description, while for a massive graviton, we introduce a novel gauge invariant action in order to show duality invariance.
\vskip 1cm


\section{INTRODUCTION}
The duality symmetry is one of the concepts most relevant in high energy physic in these last times. In particular, the electromagnetic duality has played an important role from forty years ago, since the advent of supergravity until the recent results in superstring and M theory. The aspects of linear duality symmetries are well known, but the inclusion of sources and non linear generalizations (some few cases of interacting theories are known) open new challenges.
Duality symmetry was initially understood at the level of the equations of motion \cite{GaZu}. The first successful attempt to establish electromagnetic duality symmetry at the level of the action was achieved in \cite{DeTe} after solving the Gauss constraint. The cost was the lost of explicitly Lorentz invariance of the Maxwell action, when it is expressed in terms of transverse physical variables ($\partial_i \pi_i ^T = 0 = \partial_i A_i ^T$). The Hamiltonian first order action:
\begin{equation}\label{1}
I = \int d^4 x [\pi_i ^T A_i ^T - \frac{1}{2}( \vec{\pi}^T .\vec{\pi}^T + (\nabla  \times {\vec{A}}^T )^2 ) ],
\end{equation}
is invariant under the following (spatial non-local) duality transformations
\begin{equation}\label{2}
\delta \vec{\pi}^T = \nabla \times \vec{A}^T , \quad \delta \vec{A}^T = \frac{1}{\nabla^2}\nabla \times \vec{E}^T .
\end{equation}
Since $\pi_i ^T$ satisfies the Gauss constraint: $\partial_i \pi_i ^T = 0$, it is possible to introduce a second potential : $\pi_i ^T = \epsilon_{ijk}\partial_j \tilde{A}_i ^T$ to achieve the two potentials formulation of electromagnetism \cite{ScSen}:
\begin{equation}\label{3}
I = \int d^4 x [\frac{1}{4}\epsilon^{ijk}F_{jk}^\alpha \mathcal{L}^{\alpha\beta}\dot{A} _i^\beta  - \frac{1}{2}F_{ij}^\alpha F_{ij}^\alpha ],
\end{equation}
where $A_i ^\alpha \equiv (A_i, \tilde{A}_i)$, $F_{ij}^\alpha \equiv \partial_i A_j ^\alpha - \partial_j A_i^\alpha$ (it is understood the transverse character of the potentials) and
\begin{equation}\label{4}
\mathcal{L} =
\left( \begin{array}{cc}
0 & 1 \\
-1 & 0
\end{array} \right)\ .
\end{equation}
An early approach with two potentials was proposed in \cite{Zwanzinger} where electric and magnetic sources are considered. Now, the action (\ref{3}) is invariant under local duality transformations
\begin{equation}\label{6}
A_i ^\alpha \rightarrow \mathcal{L}^{\alpha\beta} A_i^\beta .
\end{equation}
These duality transformations and the action (\ref{3}), boil down to (\ref{2}) and the Maxwell action (in the gauge $A_o = 0$) after eliminating one of the potentials using the equations of motion corresponding to the action (\ref{3}). Although the Lorentz invariance is not explicitly manifest, the action (\ref{3}) has invariance under some transformations which are equivalent to the usual Lorentz transformations on shell \cite{ScSen}.
This procedure was extended to the (non-linear) Born-Infeld electrodynamics \cite{KhouParra} and for antisymmetric fields in any dimensions \cite{BunsterHenn1}.
Despite the difficulties to implement duality invariance in Lorentz invariant action, there are two ways to circumvent this difficult, namely, the introduction of infinite fields \cite{BerIsAlv} and introducing a new gauge field, called the PST field \cite{PST}, which interact with the electromagnetic potentials in a non polynomial way, and it is equivalent to (\ref{3}) after gauge fixing the PST field.

Remarkably enough is the achievement of duality symmetry for gravity \cite{HT1} after solving the hamiltonian and momentum constraints, which characterize the canonical formulation of the linearized gravity, by introducing two (pre)potentials. The action has the following form
\begin{equation}\label{7}
I = \int dtd^3 x [\epsilon^{ijk}\epsilon^{\alpha\beta}(\partial_{abj}Z_{ak}^\alpha - \nabla^2 \partial_j Z_{bj}^\alpha )\dot{Z}_{bi}^\beta ]- \int dt H,
\end{equation}
where $Z_{ij}^\alpha \equiv (\Phi_{ij}, P_{ij})$ are the prepotentials, which are related to the spatial components of the metric ($h_{ij}$) and the canonical momentum ($\pi_{ij}$) in a non-local way \cite{HT1},  and
\begin{equation}\label{8}
H = \int d^3 x [\nabla^2 Z_{ij}^\alpha \nabla^2 Z_{ij}^\alpha + \frac{1}{2}\partial_{ij}Z_{ij}^\alpha \partial_{kl}Z_{kl}^\alpha + \partial_{ij}Z_{ij}^\alpha \nabla^2 Z^\alpha - 2\partial_{ik}Z_{jk}^\alpha \nabla^2 Z_{ij}^\alpha - \frac{1}{2}\nabla^2 Z^\alpha \nabla^2 Z^\alpha ].
\end{equation}
is the Hamiltonian and $Z^\alpha = \delta^{ij}Z_{ij}^\alpha$ are the traces of the prepotentials . This action is clearly invariant under duality transformations
\begin{equation}\label{9}
Z_{ij}^\alpha \rightarrow \mathcal{L}^{\alpha\beta} Z_{ij}^\beta .
\end{equation}
The generalization to arbitrary dimensions was achieved in \cite{BuHenn2}, based on the duality relationship between the graviton field $h_{mn}$ and the generalized Curtright field $T_{m_1 ...m_{D-3},n}$ \cite{West}, \cite{BCH}.
Afterward, the duality symmetric action for higher spin was completed in ref. \cite{DeSemi}. The essential point of this achievement is clearly expressed in this work: "The key to our derivation is the remark that since free gauge field actions are (abelian) gauge invariant, they can be uniformly expressed -after elimination of constraints- in terms of the fundamental spatial gauge-invariant symmetric transverse-traceless (TT) conjugate variables". The actions which are invariant under duality transformations for higher spin has the same structure as the electromagnetic duality invariant action (\ref{3}). We will adopt this point of view when we deal with gauge invariant models for massive theories. For instance, if in the action (\ref{7}), we consider the usual transverse-longitudinal decomposition for the (pre)potentials $Z_{ij}^\alpha (= Z_{ij}^{\alpha TT} + \partial_i Z_{j}^{\alpha T} + \partial_j Z_{i}^{\alpha T} + \partial_{ij}\alpha + \delta_{ij}\beta )$, it takes the form
\begin{equation}\label{10}
I = \int d^4 x [\frac{1}{2}\epsilon^{ijk}\partial_j \nabla^2 Z_{il}^{TT\alpha} \mathcal{L}^{\alpha\beta}\dot{Z} _{kl}^{TT\beta}  - \frac{1}{2}\nabla^2 Z_{ij}^{TT\alpha} \nabla^2 Z_{ij}^{TT\alpha} ].
\end{equation}
Moreover, the partially massless phenomenon for massive gravitons in (A)dS \cite{DN}, where the mass of the graviton is fine tuned to a value  related to the cosmological constant, exhibits an electromagnetic duality invariance \cite{DW}, where a massive graviton acquires an additional gauge invariance due to the partially massless effect. Recently, duality properties of the Horava gravity \cite{Horava}, which has in common with (\ref{7}) a higher order spatial derivative hamiltonian, was studied and discussed \cite{CorGar} in the framework of ref. \cite{BCH}.

Another example of a duality invariant action (which will be used in this paper), is the case of an antisymmetric field ($B_{mn}$) whose its field strength is $H_{mnp} \equiv \partial_m B_{np} + \partial_n B_{pm} + \partial_p B_{mn}$. This is a particular case discussed in ref. \cite{BunsterHenn1}, where the electromagnetic duality was considered for $p$ forms in D dimensions.
For an antisymmetric field, the canonical action is
\begin{equation}
I = \int d^4 x [\pi_{ij} \dot{B}_{ij} - \pi_{ij}\pi_{ij} - \frac{1}{12}H_{ijk}H_{ijk} + 2 B_{oi} \mathcal{F}_i ],
\end{equation}
where, $\pi_{ij}$ is the canonical momentum associated to $B_{ij}$ and $\mathcal{F}_i \equiv \partial_j \pi_{ij} \simeq 0$ is the corresponding Gauss constraint whose solution is $\pi_{ij} = -\frac{1}{2}\epsilon_{ijk}\partial_k\phi$. After decomposing the spatial antisymmetric field as $B_{ij} = \epsilon_{ijk} \frac{\partial_k}{\sqrt{-\nabla^2}}\psi + \partial_i b_j ^T - \partial_j b_i ^T$, the action will depend only of the dual variables ($\phi$,$\psi$):
\begin{equation}\label{19}
I = \int d^4 x [ (\sqrt{-\nabla^2}\psi \dot{\phi} - \frac{1}{2}\partial_i \phi\partial_i \phi -  \frac{1}{2}\partial_i \psi\partial_i \psi],
\end{equation}
which is invariant under: $\phi \rightarrow \psi$ and $\psi \rightarrow -\phi$.

In summary, the method for obtaining duality invariant action relies on the fact of renouncing spacetime covariance in an explicit way. This not mean breaking of Lorentz invariance. The Lorentz and duality symmetries have a closer connection \cite{BHDS} which was anticipated in \cite{ScSen} and \cite{HT88}. Starting up with a Lorentz and gauge invariant action, the hamiltonian formulation (described by a first order action) is constructed and the Gauss constraints are solved introducing new potentials. These new set of potentials (named electric and magnetic potentials) are the essential objects from which the new duality invariant action is constructed. In this paper, we generalize the duality invariant actions for massive vectorial (spin 1) and tensorial (spin 2) fields, which are described by gauge invariant actions which are obtained by dualization of the usual Stueckelberg formulations of these fields. The next section, the massive gauge invariant vectorial model is considered and we look for the duality properties in terms of the transverse and longitudinal degrees of freedom. Next, we construct a new gauge invariant action for the massive graviton from which, in the last section, we will establish the duality invariance at the level of the action. It is worth recalling that for massive theories there is no covariant duality relationship between field strengths and its dual, which interchanges field equations and Bianchi identities. Throughout this work we use $\eta_{mn}$ mostly positive and the convention of $\epsilon^{oijk} \equiv \epsilon^{ijk}$. We restrict our results to four dimensions.

\section{DUALITY FOR MASSIVE SPIN-1}
It is well known that the massive vectorial field can be described by a gauge invariant formulation which involved the
coupling of the vectorial field $A_m$ with an antisymmetric field $B_{mn}$ through a BF topological term \cite{CreSch}. The action is
\begin{equation}\label{2.1}
 I = \int d^4 x [-\frac{1}{4}F_{mn}F^{mn} - \frac{1}{12}H^{mnp}H_{mnp} - \frac{\mu}{4}\epsilon^{mnpq}B_{mn}F_{pq}],
\end{equation}
where $F_{mn} = \partial_m A_n - \partial_n A_m$ and $H_{mnp} = \partial_m B_{np} + \partial_n B_{pm} + \partial_p B_{mn}$.
This action is invariant under gauge transformations : $\delta A_m = \partial_m \lambda$ and $\delta B_{mn} = \partial_m \lambda_n - \partial_n \lambda_m$ and it is dual equivalent to the usual Stueckelberg formulation for massive spin 1 \cite{Khou}. The field equations associated with (\ref{2.1}) are
\begin{equation}
\partial_n (F^{nm} - \frac{\mu}{2}\epsilon^{mnpq}B_{pq}) = 0 \quad and \quad \partial_p (H^{pmn} - \mu\epsilon^{mnpq}A_{q}) = 0
\end{equation}
and the Bianchi identities are
\begin{equation}
\epsilon^{mnpq}\partial_n F_{pq} = 0 \quad and \quad \epsilon^{mnpq}\partial_q H_{mnp} = 0 .
\end{equation}
There is no local neither non-local duality transformations that exchange these covariant field equations with the Bianchi identities and viceversa.
The goal will be to find an action in terms of pre potentials with invariance under local duality transformations.
In order to reach a non Lorentz invariant duality action, we must start with the Hamiltonian approach. The canonical momenta are
\begin{eqnarray}\label{2.2}
\pi^i &=& \frac{\delta \textit{L}}{\delta \partial_o A_i} = F_{oi} - \frac{1}{2}\mu \epsilon_{ijk}B_{jk} {\nonumber} \\
\pi^{ij} &=& \frac{\delta \textit{L}}{\delta \partial_o B_{ij}} = \frac{1}{2}H_{oij}.
\end{eqnarray}
As usual, $A_o$ and $B_{oi}$ are multipliers. The action has the following canonical form

\begin{equation}\label{2.3}
 I = \int d^4 x [\pi^i \dot{A} _i + \pi^{ij} \dot{B} _{ij} - \mathcal{H} + A_o \verb"G" + B_{oi} \verb"G" ^i],
\end{equation}
where the Hamiltonian density is
\begin{equation}\label{2.4}
 \mathcal{H} = \frac{1}{2}(\pi_i + \frac{1}{2}\mu\epsilon_{ijk}B^{jk})(\pi^i + \frac{1}{2}\mu\epsilon^{iab}B_{ab}) + \frac{1}{4}F_{ij}F^{ij} + \pi_{ij}\pi^{ij} + \frac{1}{12}H_{ijk}H^{ijk}
\end{equation}
and the Gauss constraints are
\begin{equation}\label{2.5}
\verb"G" = \partial_i \pi^i = 0
\end{equation}
and
\begin{equation}\label{2.6}
\verb"G" ^i = \partial_j (\pi^{ij} + \frac{1}{2}\mu\epsilon^{ijk}A_{k}) = 0.
\end{equation}
These constraints are easily solved  (locally) for the momenta and lead to the "magnetic" potentials $\tilde{A}_i$ and $\phi$:
\begin{equation}\label{2.7}
\pi^i = \epsilon^{ijk}\partial_j \tilde{A}_k
\end{equation}
and
\begin{equation}\label{2.8}
\pi^{ij} + \frac{1}{2}\mu\epsilon^{ijk}A_k = \frac{1}{2}\epsilon^{ijk}\partial_k \phi .
\end{equation}
When these solutions are considered, the action is written as
\begin{eqnarray}\label{2.9}
 I = \int d^4 x &[&\frac{1}{2}\epsilon^{ijk}\mu(B_{jk} + \frac{1}{\mu}\tilde{F}_{jk})(\dot{A}_i - \frac{1}{\mu}\partial_i \dot{\phi}) - \frac{1}{2}(B_i B^i + H^2)  {\nonumber} \\
 &-& \frac{1}{2}\mu ^2 (A_i - \frac{1}{\mu}\partial_i \phi)^2 - \frac{1}{4}\mu ^2(B_{ij} + \frac{1}{\mu}\tilde{F}_{ij})^2 ] ,
\end{eqnarray}
where $\tilde{F}_{ij} = \partial_i \tilde{A}_j - \partial_j \tilde{A}_i$ and we have introduced the magnetic fields
\begin{equation}
B_i \equiv \frac{1}{2}\epsilon_{ijk}F_{jk} \quad H \equiv \frac{1}{6}\epsilon_{ijk}H_{ijk}.
\end{equation}
At this stage, we have invariance under the following gauge transformations
\begin{eqnarray}\label{2.10}
\delta A_i &=& \partial_i \lambda , \quad \delta B_{ij} = \partial_i \lambda_j - \partial_j \lambda_i  {\nonumber} \\
\delta \tilde{A}_i &=& \partial_i \tilde{\lambda} - \mu\lambda_i , \quad \delta\phi = \mu\lambda ,
\end{eqnarray}
which allow us to fix the gauges $\phi = 0$ and $\tilde{A}_i = 0$. If we consider this possibility, it is easily recognized that
$\frac{1}{2}\mu\epsilon^{ijk}B_{jk}$ as the canonical momentum associated to $A_i$ and the hamiltonian first order action for the
massive vectorial field arises. But, we need all the potentials which have emerged in order to have duality invariance in the action. To reach this goal
we decompose the fields in its irreducible transverse and longitudinal components:
\begin{equation}\label{2.11}
A_i = A_i ^T + \partial_i A^L , \quad B_{ij} = \epsilon_{ijk}\partial_k \psi + \partial_i b_j ^T - \partial_j b_i ^T.
\end{equation}
The longitudinal component of the $\tilde{A}_k$ field is absent from the beginning (see eq. \ref{2.7}),
while the longitudinal component of $A_i$ is absorbed with the scalar field $\phi$. The first term in the decomposition
for the antisymmetric field $B_{ij}$ is particular for $D = 4$; the scalar field $\psi$ will be the dual partner of $\phi$. Moreover, we define $a_i ^T \equiv \frac{1}{\mu}A_i ^T + b_i$ and its field strength $f_{ij} = \partial_i a_j - \partial_j a_i$. Then, the action is written out as (from now on the transverse and longitudinal characters are understood)
\begin{eqnarray}\label{2.12}
 I = \int d^4 x [&\frac{1}{2}&\mu\epsilon^{ijk}f_{jk}\dot{A} _i - \frac{1}{2}\mu^2 A_i A_i - \frac{1}{2}(F_{ij}F_{ij} + f_{ij}f_{ij})  {\nonumber} \\
 &+& (\nabla^2 \psi )\dot{\phi} -\frac{1}{2}\mu^2 \partial_i \psi \partial_i \psi - \frac{1}{2}\partial_i \phi \partial_i \phi - \frac{1}{2}\nabla^2 \psi \nabla^2 \psi ].
\end{eqnarray}
Thus, the pairs of dual partners: $(A_i , a_i)$ and $(\phi , \psi )$ are decoupled. Finally, we make the following definitions:
\begin{equation}
a_i = \frac{1}{\mu}\sqrt{1 - \frac{\mu^2}{\nabla^2}}\hat{A}_i , \quad \phi = \sqrt{\mu^2 - \nabla^2}\varphi ,
\end{equation}
in order to arrive to the following form of the action
\begin{eqnarray}\label{2.12}
 I = \int d^4 x &[&\frac{1}{2}\sqrt{1 - \frac{\mu^2}{\nabla^2}}\epsilon^{ijk}\hat{F}_{jk}\dot{A} _i - \frac{1}{2}\mu^2 A_i A_i - \frac{1}{2}\mu^2 \hat{A}_i \hat{A}_i
 - \frac{1}{2}(F_{ij}F_{ij} + \hat{F}_{ij}\hat{F}_{ij})  {\nonumber} \\
 &+& (\nabla^2 \psi )\sqrt{\mu^2 - \nabla^2}\dot{\varphi} -\frac{1}{2}\mu^2 \partial_i \psi \partial_i \psi - \frac{1}{2}\mu^2 \partial_i \varphi \partial_i \varphi - \frac{1}{2}\nabla^2 \varphi \nabla^2 \varphi - \frac{1}{2}\nabla^2 \psi \nabla^2 \psi ],
\end{eqnarray}
or
\begin{eqnarray}\label{2.13}
 I = \int d^4 x &[&\frac{1}{4}\sqrt{1 - \frac{\mu^2}{\nabla^2}}\epsilon^{ijk}F_{jk}^\beta \mathcal{L}^{\alpha\beta}\dot{A} _i^\alpha - \frac{1}{2}\mu^2 A_i^\alpha A_i^\alpha - \frac{1}{2}F_{ij}^\alpha F_{ij}^\alpha   {\nonumber} \\
 &+& \frac{1}{2}(\nabla^2 \Phi^\beta )\sqrt{\mu^2 - \nabla^2}\mathcal{L}^{\alpha\beta}\dot{\Phi^\alpha} - \frac{1}{2}\mu^2 \partial_i \Phi^\alpha \partial_i \Phi^\alpha -  \frac{1}{2}\nabla^2 \Phi^\alpha \nabla^2 \Phi^\alpha ],
\end{eqnarray}
where we have introduced the conventional notation: $A_i ^\alpha = (A_i , \hat{A}_i ) , \quad  \Phi^\alpha = (\varphi , \psi)$.
The action is clearly invariant under duality transformations
\begin{equation}\label{2.15}
A_i ^\alpha \rightarrow \mathcal{L}^{\alpha\beta} A_i^\beta \quad and \quad \Phi ^\alpha \rightarrow \mathcal{L}^{\alpha\beta}\Phi^\beta ,
\end{equation}

The field equations obtained from (\ref{2.13}) are
\begin{equation}
\sqrt{1 - \frac{\mu^2}{\nabla^2}}\epsilon^{ijk}\mathcal{L}^{\alpha\beta}\partial_j \dot{A} _i^\beta + (\mu^2 - \nabla^2 )A_i ^\alpha = 0
\end{equation}
and
\begin{equation}
\sqrt{\mu^2 - \nabla^2}\mathcal{L}^{\alpha\beta}\dot{\Phi}^\beta - (\mu^2 - \nabla^2 )\Phi^\alpha = 0 .
\end{equation}
By iteration of these first order differential equations, it is easily checked that the potentials $A_i ^\alpha$ and $\Phi^\alpha$ satisfy the Klein-Gordon equation: \begin{equation}
(\Box - \mu^2 ) A_i ^\alpha = 0 \quad and \quad (\Box - \mu^2 ) \Phi^\alpha = 0 .
\end{equation}

The canonical momenta are slightly modified by the presence of a mass term and the following second class constraints are derived
\begin{equation}\label{2.16}
\Upsilon_i ^\alpha \equiv \pi_i ^\alpha - \frac{1}{4}\sqrt{1 - \frac{\mu^2}{\nabla^2}}\epsilon^{ijk}F_{jk}^\beta \mathcal{L}^{\alpha\beta} \approx 0 ,\quad
\Gamma^\alpha \equiv \pi^\alpha - \frac{1}{2}\sqrt{\mu^2 - \nabla^2}(\nabla^2 \Phi^\beta ) \mathcal{L}^{\alpha\beta} \approx 0 ,
\end{equation}
from which the Poisson bracket for these second class constraints now read as
\begin{equation}\label{2.17}
[\Upsilon_{i(x)}^\alpha , \Upsilon_{j(y)}^\beta ] =  \sqrt{1 - \frac{\mu^2}{\nabla^2}} \mathcal{L}^{\alpha\beta}\epsilon_{ijk}\partial_k \delta_{(x - y)}, \quad
[\Gamma_{(x)}^\alpha , \Gamma_{(y)}^\beta ] =  \sqrt{\mu^2 - \nabla^2} \mathcal{L}^{\alpha\beta} \nabla^2 \delta_{(x - y)} .
\end{equation}

These momenta must transform under duality as $\pi_i ^\alpha \rightarrow \mathcal{L}^{\alpha\beta} \pi_i^\beta$ and $\pi ^\alpha \rightarrow \mathcal{L}^{\alpha\beta}\Phi^\pi$ in order to keep the invariance of the Poisson bracket under duality.
The action (\ref{2.13}) is invariant under the following global transformations
\begin{equation}
\delta A_i ^\alpha = x^o v^j \partial_j A_i ^\alpha + \sqrt{1 - \frac{\mu^2}{\nabla^2}}(\vec{v}.\vec{x})\epsilon^{ijk}\mathcal{L}^{\alpha\beta}\partial_j A_k ^\beta
\end{equation}
and
\begin{equation}
\delta \Phi^\alpha  = x^o v^j \partial_j \Phi^\alpha + \sqrt{\mu^2 - \nabla^2}(\vec{v}.\vec{x})\mathcal{L}^{\alpha\beta}\nabla^2 \Phi ^\beta ,
\end{equation}
where $\vec{v}$ is an arbitrary constant three dimensional vector. If we use the solutions of the constraints (\ref{2.16}), these become the usual Lorentz boosts transformations
\begin{equation}
\delta \mathbf{X}  = x^o v^j \partial_j \mathbf{X} + (\vec{v}.\vec{x})\partial_o \mathbf{X} ,
\end{equation}
where $\mathbf{X} = (A_i ^1 , \phi )$ and we have considered that $2\pi_i ^1 = \partial_o A_i ^1$ and $2\pi ^1 = \partial_o \phi$. Moreover the action (\ref{2.13}) is manifestly invariant under rotations.

In the limit $\mu \rightarrow 0$, the action (\ref{2.13}) is the sum of the actions (\ref{3}) and (\ref{19}) which indicate that there is no discontinuity of the number of degrees of freedom as can be seen from the covariant Lorentz action (\ref{2.1}), where the scalar field, which provides mass to the vectorial field and  represented by the antisymmetric field $B_{mn}$ is cleanly decoupled from the vectorial action in this limit.

\section{GAUGE INVARIANT FORMULATION FOR THE MASSIVE GRAVITY}

For massive gravity, we need a new gauge invariant formulation for the symmetric field ($h_{mn}$), which generalizes the action (\ref{2.1}) for massive spin 2. To construct this action, we start with the usual Stueckelberg formulation for the Fierz-Pauli action :
\begin{equation}\label{3.1}
 I = I_{lin} + \int d^4 x [- \frac{1}{4}\mu^2 (h_{mn} + \partial_m a_n + \partial_n a_m )^2 + \frac{1}{4}\mu^2 (h + 2\partial_m a_m )^2] ,
\end{equation}
where
\begin{equation}\label{3.2}
 I_{lin} = \int d^4 x [-\frac{1}{4}\partial_p h_{mn}\partial_p h^{mn} + \frac{1}{4}\partial_p h \partial_p h + \frac{1}{2}\partial_n h_{mn}\partial_p h^{mp} - \frac{1}{2}\partial_m h\partial_n h^{mn}],
\end{equation}
is the linearized Einstein action and $a_m$ is the vectorial Stuckelberg field which guarantee gauge invariance under the following transformations:
\begin{equation}\label{3.3}
\delta h_{mn} = \partial_m \xi_n + \partial_n \xi_m , \quad \delta a_m = -\xi_m .
\end{equation}
We can dualize this action. For this process, we substitute $\partial_m a_n + \partial_n a_m $ by a symmetric tensor $g_{mn}$ and introducing a new term into the action which enforces that linearized curvature tensor $R_{mnpq(g)} (\equiv \partial_{np}g_{mq} + \partial_{mq}g_{np} - \partial_{nq}g_{mp} - \partial_{mp}g_{nq})$ vanishes through a Lagrange multiplier $B_{mnpq}$ and which has the same symmetries as the linearized curvature tensor $R_{mnpq}$.
Then, we have the following action
\begin{equation}\label{3.4}
 I = I_{lin} + \int d^4 x [- \frac{1}{4}\mu^2 (h_{mn} + g_{mn})(h^{mn} + g^{mn}) + \frac{1}{4}\mu^2 (h + g)^2 + \frac{1}{8}B^{mnpq}R_{mnpq(g)}] .
\end{equation}
After considering the constraint imposed by the multiplier $B_{mnpq}$, which tell us that $g_{mn} = \partial_m a_n + \partial_n a_m$, we obtain (\ref{3.1}). On the other hand, we can determine $g_{mn}$ by using its field equation
\begin{equation}\label{3.5}
g_{mn} = -h_{mn} - \frac{1}{\mu^2}(\partial_p \partial_q B_{mpnq} - \frac{1}{3}\eta_{mn}\partial_p \partial_q B_{rprq}) \equiv -h_{mn} + f_{mn} ,
\end{equation}
where we have defined $f_{mn} = - \frac{1}{\mu^2}(\partial_p \partial_q B_{mpnq} - \frac{1}{3}\eta_{mn}\partial_p \partial_q B_{rprq})$. 
After introducing (\ref{3.5}) into the action (\ref{3.4}), we reach (and redefining $B_{mnpq} \rightarrow \mu B_{mnpq}$) the following result
\begin{eqnarray}\label{3.10}
 I = \int d^4 x [ &-& \frac{1}{4}\partial_p h_{mn}\partial_p h^{mn} + \frac{1}{4}\partial_p h \partial_p h + \frac{1}{2}\partial_n h_{mn}\partial_p h^{mp}  - \frac{1}{2}h\partial_m \partial_n h^{mn} ] - \frac{1}{8}\mu\int d^4 x B^{mnpq}R_{mnpq(h)}  {\nonumber} \\
 &-& \frac{1}{8}\mu\int d^4 x B^{mnpq}R_{mnpq(f)} - \frac{1}{4}\mu^2 \int d^4 x (f_{mn}f^{mn} - f^2 )
\end{eqnarray}
and this action expressed only en terms of the fields $h_{mn}$ and $B_{mnpq}$ is written as
\begin{equation}\label{3.68}
 I = I_{lin} +I_{DST} - \frac{\mu}{8}\int d^4 x B^{mnpq}R_{mnpq(h)},
\end{equation}
where
\begin{equation}\label{3.8}
 I_{DST} = \frac{1}{4}\int d^4 x [\partial_p \partial_q B_{mpnq}\partial_r \partial_s B_{mrns} - \frac{1}{3} (\partial_p \partial_q B_{npnq})^2 ],
\end{equation}
is the free ghost higher derivative action found by Deser, Siegel and Townsend \cite{DST} and which is an alternative description of the Maxwell action. In the appendix, some aspects of this action are reviewed. 
The action (\ref{3.68}) is invariant under the following gauge transformations:
\begin{equation}\label{3.9}
\delta h_{mn} = \partial_m \xi_n + \partial_n \xi_m , \quad \delta B^{mnpq} = \epsilon^{mnrs}\partial_r \lambda^{pq},_s + (mn) \leftrightarrow (pq) ,
\end{equation}
where the gauge parameters are antisymmetric in the pair $(pq)$ i.e.  $\lambda^{pq},_s = -\lambda^{qp},_s$ without any additional properties in its indices. The action (\ref{3.8}) has a conformal invariance which here is lost by the coupling term $\sim B^{mnpq}R_{mnpq(h)}$.

The action (\ref{3.10}) is our gauge invariant formulation for massive spin 2 where we consider $h_{mn}$, $B_{mnpq}$ and $f_{mn}$ as independent fields. 
The field $f_{mn}$ is an auxiliary field that when its equation of motion is used, the action (\ref{3.8}) is obtained. This way of describing the field $B^{mnpq}$ with the use of an auxiliary field $f^{mn}$ was introduced in \cite{DST}.
The coefficients in this action are the same for any dimensions, while the coefficient $\frac{1}{3}$ in the second term of (\ref{3.8}) must be replaced by $\frac{1}{D - 1}$. In particular, in three dimensions the action (\ref{3.10}) becomes the action for the new massive gravity \cite{BHT}. In fact, in three dimensions, we have the following identity: $R_{mnpq} = \epsilon_{mnr}\epsilon_{pqs}G^{rs}$ which tell us that the curvature is completely determined by the
Einstein tensor $G_{mn}$ since the conformal tensor vanishes identically. Then, we rewrite the term $- \frac{1}{8}B^{mnpq}R_{mnpq(h + f)}$ as $-\frac{1}{2}B^{mn}G_{mn(h + f)}$ where we have defined $B^{mn} \equiv \epsilon^{mpq}\epsilon^{nrs}R_{pqrs}$ and the action can be rewritten as
\begin{equation}\label{3.12}
I_{3D} = \int d^3 x [\frac{1}{4}h^{mn}G_{mn} - \frac{1}{2}B^{mn}G_{mn(h + f)} - \frac{1}{2}\mu^2 (f^{mn}f_{mn} - f^2) ],
\end{equation}
which becomes the action for the new massive gravity after eliminating the $h_{mn}$ field through its field equation.

For sake of completeness, let us sketch the dynamical content of the action(\ref{3.10}) to confirm that our model propagates the five degrees of freedom of the massive spin 2 in four dimensions. The temporal and spatial components are denoted by $h_{mn}: (h_{oo}\equiv\psi ,h_{oi},h_{ij})$, $f_{mn}: (f_{oo}\equiv \theta ,f_{oi},f_{ij})$ and ($S_{ij}\equiv B_{iooj} = S_{ji}, W_{ij}\equiv \partial_k (B_{oikj} + B_{ojki}) = W_{ji}, V_{ij}\equiv \partial_{kl}B_{iklj} = V_{ji}$). When the action is written out in terms of these components, the $B_{ijkl}$ variable appears as a Lagrange multiplier associated with the constraint $R_{ijkl(h + f)} = 0$, which can be solved (locally) as $f_{ij} = h_{ij} + \partial_i \kappa_j + \partial_j \kappa_i$. We can use the gauge invariance (\ref{3.9}) in order to fix the gauge $\kappa_i = 0$. Now, we consider the usual transverse and longitudinal decompositions:
\begin{equation}\label{3.13}
h_{ij} = \chi_{ij} + \partial_i h_j + \partial_j h_i + \partial_{ij}\sigma + \delta_{ij}\tau , \quad h_{oj} = u_{i} + \partial_i v ,
\end{equation}

\begin{equation}\label{3.15}
f_{ij} = \tau_{ij} + \partial_i f_j + \partial_j f_i + \partial_{ij}\lambda + \delta_{ij}\alpha , \quad f_{oj} = r_{i} + \partial_i t ,
\end{equation}
and
\begin{equation}\label{3.17}
S_{ij} = s_{ij} + \partial_i s_j + \partial_j s_i + \partial_{ij}s + \delta_{ij}\beta , \quad  W_{ij} = w_{ij} + \partial_i w_j + \partial_j w_i + \partial_{ij}w + \delta_{ij}\gamma
\end{equation}

where $\chi_{ij} , \tau_{ij} , s_{ij}$ and $w_{ij}$ are transverse and traceless tensor, $h_j , u_{i} , f_i , r_{i} , s_i$ and $w_i$ are transverse vectors, while $\sigma , \tau , v, \lambda , \alpha, t, s, \beta , w$ and $\gamma$ are scalars. The solution of the constraint imposed by the $B_{ijkl}$ tell us
\begin{equation}
\tau_{ij} = - \chi_{ij}, \quad f_i = - h_i , \quad \alpha = - \tau , \quad \lambda = - \sigma .
\end{equation}
It is easily checked that $s_{ij}$ and $w_{ij}$ do not enter into the action, then the transverse and traceless sector of the action is, as it must be:
\begin{equation}\label{3.20}
\frac{1}{4} \chi_{ij} \Box \chi_{ij} - \frac{1}{4}\mu^2 \chi_{ij}\chi_{ij},
\end{equation}
which represent the propagation of the two degrees of freedom of the tensorial sector
The vectorial sector of the action is
\begin{equation}\label{3.21}
- u_i \nabla^2 u_i + 2u_i \nabla^2 \dot{h}_i - \dot{h}_i \nabla^2 \dot{h}_i + 2\mu u_i \nabla^2 \dot{s}_i + 2\mu u_i \nabla^2 w_i +
2\mu r_i \nabla^2 \dot{s}_i + 2\mu r_i \nabla^2 w_i + \mu^2 r_i r_i + \mu^2 h_i \nabla^2 h_i
\end{equation}
and it is clear that $u_i$ and $r_i$ are auxiliary fields that are determined using its field equations. Its values are
\begin{equation}\label{3.22}
u_i = \dot{h}_i + \mu \dot{s}_i + \mu w_i , \quad r_i = - \frac{\nabla^2}{\mu}\dot{s}_i - \frac{\nabla^2}{\mu}w_i
\end{equation}
and after substituting (\ref{3.22}) into the vectorial sector of the action, $w_�$ arises as un auxiliary field and can be determined through its
field equation
\begin{equation}\label{3.23}
w_i = - \mu \dot{s}_i - \frac{\mu}{\mu^2 - \nabla^2}\dot{h}_i .
\end{equation}
We can see that after taking into account (\ref{3.23}), the vectorial sector of the action depends on $h_i$ only and redefining $h_i = \frac{1}{\mu}\sqrt{1 - \frac{\mu^2}{\nabla^2}}\bar{h}_i$ we end with the propagation of the two degrees of freedom of the vectorial sector
\begin{equation}\label{3.24}
\sim \bar{h}_i \Box \bar{h}_i - \mu^2 \bar{h}_i \bar{h}_i .
\end{equation}
In the scalar sector, we found that $\theta , \gamma , t$ and $w$ are not dynamical variables (these are multipliers or are solved) and this sector reduces to
the scalar sector of the Fierz-Pauli action:
\begin{equation}\label{3.25}
2\psi \nabla^2 \tau + 4v\nabla^2 \dot{\tau} + 4\sigma\nabla^2 \ddot{\tau} + 3\tau \ddot{\tau} - \tau \nabla^2 \tau
+2\mu^2 \sigma \nabla^2 \tau + 3\mu^2 \tau \nabla^2 \tau - \mu^2 v\nabla^2 v - \mu^2 \psi \nabla^2 \sigma - 3\mu^2 \psi\tau .
\end{equation}
Clearly, $\psi$ is a multiplier and its corresponding constraint is $2\nabla^2 \tau - \mu^2 \nabla^2 \sigma - 3\mu^2 \tau = 0$ which is solved for $\sigma$ (=$\frac{2}{\mu^2}\tau - \frac{3}{\nabla^2}\tau$), and after introducing this value into the scalar sector, the $v$ variable appears quadratically without temporal derivatives and its field equation determines it : $v = \frac{2}{\mu^2}\dot{\tau}$. With this result at hand, we reach the final unconstrained form of the propagation of one scalar degree of freedom :
\begin{equation}
\sim \tau (\Box  - \mu^2 )\tau .
\end{equation}
Thus, we have seen that our action for describing in a gauge invariant way the massive graviton, propagates the usual five free ghost degrees of freedom.

\section{DUALITY FOR MASSIVE GRAVITY}
In this section we show that the unconstrained hamiltonian form of the gauge invariant action (\ref{3.10}) admits the introduction of new potentials in order
to reach a duality invariance for the massive spin 2. To begin with, we need a first order canonical action of (\ref{3.10}). The canonical momenta are
\begin{equation}\label{4.1}
\pi_{ij} =\frac{\delta I}{\delta\dot{h}_{ij}} = \frac{1}{2}\dot{h}_{ij} - \frac{1}{2}\delta_{ij}\dot{h} - \frac{1}{2}(\partial_i h_{oj} + \partial_j h_{oi}) + \delta_{ij}\partial_k h_{ok} + \frac{1}{2}\mu b_{ij} ,
\end{equation}
where $h$ is tha spatial trace of $h_{ij}$ and we have defined $b_{ij} = \dot{S}_{ij} + W_{ij}$,
\begin{equation}\label{4.2}
\pi_i =\frac{\delta I}{\delta\dot{h}_{oi}} = 0, \quad \sigma_{ij} = \frac{\delta I}{\delta\dot{f}_{ij}} = \frac{1}{2}\mu b_{ij}, \sigma_i = \frac{\delta I}{\delta\dot{f}_{oi}} = 0 .
\end{equation}
The first order canonical action is
\begin{equation}\label{4.3}
I \int d^4 x [\pi_{ij}\dot{h}_{ij} + \frac{1}{2}\mu b_{ij}\dot{f}_{ij} - \mathbb{H} - n_i \mathcal{H}_i - n \mathcal{H}] ,
\end{equation}
where $n_i$ and $n$ are the shift and lapse functions associated with the momentum ($\mathcal{H}_i \approx 0$) and Hamiltonian ($\mathcal{H}_i \approx 0$) constraints, respectively.
\begin{equation}\label{4.4}
\mathcal{H}_i \equiv \partial_j \pi_{ij} \approx 0 , \quad \mathcal{H} \equiv \nabla^2 h - \partial_{ij}h_{ij} - \mu \partial_{ij}S_{ij} \approx 0 .
\end{equation}
The Hamiltonian constraint is modified by the presence of the mass term in the action, while the momentum constraint remains intact.

The Hamiltonian density is
\begin{eqnarray}\label{4.5}
\mathbb{H} &=& [\pi_{ij} - \frac{1}{2}\mu b_{ij}][\pi^{ij} - \frac{1}{2}\mu b^{ij}]  - [\pi - \frac{1}{2}\mu b]^2 + \mathcal{R}  + \frac{1}{2}\mu (h_{ij} + f_{ij})V_{ij} {\nonumber} \\
&-& \mu (\partial_j b_{ij})f_{oj} - \frac{1}{2}\mu^2 f_{oi}f_{oi} + \frac{1}{4}\mu^2 f_{ij}f_{ij} - \frac{1}{4}\mu^2 f_{ii}f_{jj} + \frac{1}{2}f_{oo} (\mu S_{ij} + \mu^2 f_{ii}),
\end{eqnarray}
being
\begin{equation}\label{4.6}
\mathcal{R} \equiv \frac{1}{4}\partial_k h_{ij}\partial_k h^{ij} - \frac{1}{4}\partial_k h \partial_k h - \frac{1}{2}\partial_k h_{ki}\partial_j h^{ji} + \frac{1}{2}\partial_i h \partial_j h^{ji} .
\end{equation}
Note that $f_{oo}$ and  $B_{ijkl}$ are Lagrange multipliers associated with the constraints: $\partial_{ij}S_{ij} + \mu f_{ii} = 0$ and $R_{ijkl(h + f)} = 0$. The former allow us express the double spatial derivatives of $S_{ij}$ in terms of the spatial trace of $f_{ij}$, while the latter tell us that: $f_{ij} = - h_{ij} + \partial_i a_j + \partial_j a_i$ and since we have gauge invariance one can set $a_i = 0$. Also, $f_{oi}$ is an auxiliary field and its field equation allows determine it :
$f_{oi} = -\frac{1}{\mu}\partial_j b_{ij}$ and considering this value, the term $-\frac{1}{2}\partial_j b_{ij} \partial_k b_{ik}$ will emerge in the action.
Now, we proceed to resolve the constraints by introducing the pre potentials \cite{HT1}. The solution for the momentum constraint ($\mathcal{H}_i \approx 0$) is the same as the massless case
\begin{equation}\label{4.7}
\pi_{ij} = \epsilon_{ika}\epsilon_{jlb}\partial_{kl}\tilde{P}_{ab},
\end{equation}
where $\tilde{P}_{ij}$ is the symmetric pre potential associated with the momentum constraint. In order to solve the Hamiltonian constraint ($\mathcal{H} \approx 0$), we decompose the six components of the spatial metric in its traceless part $\hat{h}_{ij}$, (five components) and its trace $h$ (one component) as
\begin{equation}\label{4.8}
h_{ij} = \hat{h}_{ij} + \frac{1}{3}\delta_{ij}h.
\end{equation}
After considering this decomposition in the hamiltonian constraint, we see that the traceless part must satisfies the following differential equation:
\begin{equation}\label{4.9}
\partial_{ij}\hat{h}_{ij} - \frac{2}{3}\nabla^2 h + \mu^2 h = 0 .
\end{equation}
The solution is
\begin{equation}\label{4.10}
\hat{h}_{ij} = J_{ij} + (\frac{\partial_{ij}}{\nabla^2} - \frac{1}{3}\delta_{ij})(1 - \frac{3\mu^2}{\nabla^2})h,
\end{equation}
where (as introduced in \cite{HT1} )
\begin{equation}\label{4.11}
J_{ij} = \epsilon_{ikl}\partial_k \Phi_{lj} + \epsilon_{jkl}\partial_k \Phi_{li},
\end{equation}
being $\Phi_{ij}$ the symmetric pre potential linked to the hamiltonian constraint. Essentially, these pre potentials will describe the transverse and traceless spin 2 sector of our massive model like the massless case. Before considering these pre potential into the action, we redefine $\tilde{P}_{ij}$:
\begin{equation}\label{4.12}
\tilde{P}_{ij} = P_{ij} - \frac{1}{2}\frac{\mu}{\nabla^2}b_{ij} + \frac{1}{2}\frac{\mu}{\nabla^2}\delta_{ij}b .
\end{equation}
With these results, the first order canonical action is written out as
\begin{eqnarray}\label{4.13}
 I = \int d^3 x &dt& [(\pi_{ij(P)} - \frac{1}{2}\frac{\mu}{\nabla^2}(\partial_{ik}b_{jk} + \partial_{jk}b_{ik}) + \frac{1}{2}\frac{\mu}{\nabla^2}\delta_{ij}\partial_{kl}b_{kl}) \dot{J}_{ij(\Phi )} - \mathcal{R}_{(\Phi ,h)} + r_{(\Phi)} {\nonumber} \\
 &-& \pi_{ij(P)}\pi_{ij(P)} + \frac{1}{2}\pi_{(P)}^2 - \frac{1}{2}(1 - \frac{\mu^2}{\nabla^2})(\partial_j b_{ij})(\partial_k b_{ik})],
\end{eqnarray}
where
\begin{equation}\label{4.14}
\mathcal{R}_{(\Phi ,h)} = -\frac{1}{4}J_{ij}\nabla^2 J_{ij} + \frac{1}{8}\frac{\mu^4}{\nabla^2}h^2
\end{equation}
and
\begin{equation}\label{4.15}
r_{(\Phi)} = - \mu^2 \partial_k \Phi_{ij}\partial_k \Phi_{ij} + \frac{3}{2} \mu^2 \partial_j \Phi_{ij} \partial_k \Phi_{ik} + \mu^2 \Phi \partial_{ij} \Phi_{ij} - \frac{1}{2}\mu^2 \Phi\nabla^2 \Phi .
\end{equation}
Now, we consider the different irreducibles pieces of the involved variables, e.g. $\Phi_{ij} = \Phi_{ij}^{TT} + \partial_i \phi_j ^T + \partial_j \phi_i ^T + \partial_{ij}\phi + \delta_{ij}\rho$, and for $P_{ij} = P_{ij}^{TT} + \partial_i p_j ^T + \partial_j p_i ^T + \partial_{ij}\sigma + \delta_{ij}\tau$. The superscript $TT$ and $T$ mark the transverse-traceless and transverse characters of the variable, respectively. Note that $b_{ij}$ appears only as a vector: $\partial_j b_{ij}$ which we named $b_i$ ($=b_i ^T + \partial_i b$).
The total action is written out as
\begin{equation}\label{4.16a}
I = I_2 ^{TT} + I_1 ^{T} + I_0 .
\end{equation}

The transverse-traceless sector of the action boils down to (after making $(P, \Phi) \rightarrow \frac{1}{\sqrt{-\nabla^2}}(P, \Phi)$)
\begin{equation}\label{4.16}
I_2 ^{TT} = \int d^3 x dt [ \epsilon_{ijk}(\partial_j P_{kl}^{TT})\dot{\Phi}_{il}^{TT} - \frac{1}{2}\partial_k \Phi_{ij}^{TT}\partial_k \Phi_{ij}^{TT} - \frac{1}{2}\partial_k P_{ij}^{TT}\partial_k P_{ij}^{TT} - \frac{1}{2}\mu^2\Phi_{ij}^{TT}\Phi_{ij}^{TT} ]
\end{equation}
and redefining $P_{ij}^{TT} = \sqrt{1 - \frac{\mu^2}{\nabla^2}}\hat{P}_{ij}^{TT}$, we obtain
\begin{equation}\label{4.17}
I_2 ^{TT} = \int d^3 x dt [ \sqrt{1 - \frac{\mu^2}{\nabla^2}}\epsilon_{ijk}(\partial_j \hat{P}_{kl}^{TT})\dot{\Phi}_{il}^{TT} - \frac{1}{2}\partial_k \Phi_{ij}^{TT}\partial_k \Phi_{ij}^{TT} - \frac{1}{2}\partial_k P_{ij}^{TT}\partial_k P_{ij}^{TT} - \frac{1}{2}\mu^2\Phi_{ij}^{TT}\Phi_{ij}^{TT} - \frac{1}{2}\mu^2\hat{P}_{ij}^{TT}\hat{P}_{ij}^{TT} ],
\end{equation}
which is clearly invariant under duality transformation
\begin{equation}\label{4.18}
\Phi_{ij}^{TT} \rightarrow  \hat{P}_{ij}^{TT} \quad and \quad \hat{P}_{ij}^{TT} \rightarrow - \Phi_{ij}^{TT}.
\end{equation}
Next, we move on the vectorial sector, which is described by
\begin{equation}\label{4.19}
I_1 ^{T} = \int d^3 x dt [ \mu b_i ^T \epsilon_{ijk}\partial_j \dot{\phi}_k ^T - \frac{1}{2}b_i ^T (1 - \frac{\mu^2}{\nabla^2})b_i ^T  - \frac{1}{2}\mu^2\nabla^2 \phi_{i}^{T}\nabla^2 \phi_{ij}^{T}],
\end{equation}
then, after making some redefinitions on $b_i ^T$ and $\phi_i ^T$,
we reach the following form for the transverse vectorial sector of our action
\begin{equation}\label{4.20}
I_1 ^{T} = \int d^3 x dt [ \sqrt{1 - \frac{\mu^2}{\nabla^2}}\epsilon_{ijk}(\partial_j b_{k}^{T})\dot{\phi}_{i}^{T} - \frac{1}{2}\partial_k b_{i}^{T}\partial_k b_{i}^{T} - \frac{1}{2}\partial_k \phi_{i}^{T}\partial_k \phi_{i}^{T} - \frac{1}{2}\mu^2\phi_{i}^{T}\phi_{i}^{T} - \frac{1}{2}\mu^2b_{i}^{T}b_{i}^{T} ],
\end{equation}
which is clearly invariant under duality transformations
\begin{equation}\label{4.21}
\phi_{i}^{T} \rightarrow  b_{i}^{T} \quad and \quad b_{i}^{T} \rightarrow - \phi_{i}^{T}.
\end{equation}
Finally, we have in principle, three scalar variables: $\tau$ (from the trace of $\pi_{ij(P)}$), $b$ and $h$ in the scalar sector of the action
\begin{equation}\label{4.22}
I_0 = \int d^3 x dt [ \mu^2\tau \dot{h} - \mu \dot{h}(\frac{1}{2} - \frac{\mu^2}{\nabla^2})b + \frac{3}{8}\frac{\mu^4}{\nabla^2}h(1 - \frac{\mu^2}{\nabla^2})h - \mu (\nabla^2 b) \tau + \frac{1}{2}b (\nabla^2 - \frac{5}{4}\mu^2 )b ,
\end{equation}
then we make the following (canonical) identification : $\mu^2\tau - \mu (\frac{1}{2} - \frac{\mu^2}{\nabla^2})b \equiv \frac{3}{8}\frac{\mu^3}{\nabla^2}b$ in order to write the scalar action as (redefining $h \rightarrow \frac{\nabla^2}{\mu}h$)
\begin{equation}\label{4.23}
I_0 = \frac{3\mu^2}{4}\int d^3 x dt [ b\dot{h} - \frac{1}{2}h (\mu^2 - \nabla^2 )h - \frac{1}{2}\mu^2 b^2 ]
\end{equation}
and defining: $b \equiv \sqrt{1 - \frac{\nabla^2}{\mu^2}}\tilde{b}$, we arrive to
\begin{equation}\label{4.24}
I_0 = \frac{3\mu^2}{4}\int d^3 x dt [\sqrt{\mu^2 - \nabla^2}\tilde{b}\dot{h} - \frac{1}{2}h (\mu^2 - \nabla^2 )h - \frac{1}{2} \tilde{b} (\mu^2 - \nabla^2 )\tilde{b}],
\end{equation}
clearly invariant under duality transformations:
\begin{equation}\label{4.25}
h \rightarrow  \tilde{b} \quad and \quad \tilde{b} \rightarrow - h.
\end{equation}
We have established duality invariance of massive gravity through the three levels given in the action (\ref{4.16a}). The first level is achieved with the same dual partners as the massless case: $\Phi_{ij}$ and $P_{ij}$, while in the other two layers, the dual partners involved the lower spin components of $h_{mn}$ and $B_{mnpq}$. In the massless limit, our result results tend to the decoupling of the three terms with the same structure as (\ref{3}), (\ref{10}) and (\ref{19}).

\section{CONCLUSIONS}
We have seen that if we deal with adequate gauge invariant formulations for massive theories, it is possible to reach unconstrained duality invariant actions by using the method developed in \cite{DeTe}, \cite{HT1} and \cite{BunsterHenn1}; to wit: start with a hamiltonian gauge invariant first order action and solve the gauge constraints by introducing the named magnetic potentials, that together with the initial electric potentials, are the basic entities to establish duality invariance. For the case of massive theories, the actions involved new fields and the mixing with the original field are extensions of usual BF coupling for the vectorial model. We have explicitly constructed this kind of action for massive gravitons, (\ref{3.10}). One must expect that it is not possible to have a successful coupling with a gravitational background, because the action of Deser, Siegel and Townsend has serious problem with this kind of coupling \cite{DST}. But, one can attempt to reach a non linear version of (\ref{3.10}) like the Freedman-Townsend theory \cite{FredTown}. It would be interesting to make an exhaustive analysis of the action (\ref{3.10}), in particular its properties in the massless limit and when a source is included.

\section{ACKNOWLEDGMENTS}
We would have liked to express our gratitude to our institutions for financial support, but in the last years we have seen as the budget for scientific activities in the Venezuelan Universities has been reduced drastically, due (in our opinion) for mistakes and misleading policies of the Venezuelan Government. But, despite the adversities, we would like to thank to our lovely alma mater: La Universidad. We would like to thank Pio Arias and the referee for useful comments.

\appendix
\section{THE ACTION OF DESER-SIEGEL-TOWNSEND}

The action of Deser-Siegel-Townsend (\ref{3.8}) is rewritten out as
\begin{equation}\label{A1}
 I_{DST} = \frac{1}{4}\int d^4 x [ W_{mn}W^{mn}- \frac{1}{3} W^2 ],
\end{equation}
where $W_{mn} \equiv -\partial_{pq}B_{mpnq} = W_{nm}$ and $W$ is its trace. In this form, this action has the same form as introduced in \cite{DT} where
$W_{mn}$ is considered an independent field and subject to the constraint: $\partial_n W^{mn} = 0$. If we take into account this constraint we must attempt to consider the following action
\begin{equation}\label{A2}
 I = \frac{1}{4}\int d^4 x [ W_{mn}W^{mn}- \frac{1}{3} W^2 - W^{mn}(\partial_m a_n + \partial_n a_m) ].
\end{equation}
Here $W_{mn}$ and $a_m$ are independent fields. This action was obtained by dimensional reduction in \cite{KMU}. If one looks the $a_m$ field as a multiplier, the associated constraint is just $\partial_n W^{mn} = 0$, whose local solution introduces the $B^{mpqn}$ field ($W^{mn} = \partial_{pq}B^{mpqn}$) and the fourth order action (\ref{A1}) is obtained. On the other hand, independent variations over $W^{mn}$ lead to determine it: $W^{mn} = - (\partial_m a_n + \partial_n a_m) + 2\eta_{mn}(\partial . a)$ and substituting back into \ref{A2}, the Maxwell action arises ($I_{maxwell} \sim -(\partial_m a_n - \partial_n a_m)^2$). The inclusion of a mass term into (\ref{A2}) was achieved in \cite{Dal}.

Originally this equivalence was shown through the following second order action (\ref{3.8})
\begin{equation}
I = \int d^4 x [\frac{1}{8}\int d^4 x B^{mnpq}R_{mnpq(f)} - \frac{1}{4}( f_{mn}f^{mn} - f^2 )].
\end{equation}

The fourth order action (\ref{A1}) has the following gauge transformations
\begin{equation}\label{A4}
a) \quad \delta B^{mnpq} = \epsilon^{mnrs}\partial_r \lambda^{pq},_s + (mn) \leftrightarrow (pq)
\end{equation}
and a conformal transformation
\begin{equation}
b) \quad \delta B^{mnpq} = (\eta_{mp}\eta_{nq} - \eta_{np}\eta_{mq})\varrho .
\end{equation}
This last invariance is not present in our gauge invariant description for massive spin 2, due to the coupling term (\ref{3.7}).

If the gauge parameters $\lambda_{mn,p}$ is decomposed in irreducible pieces as
\begin{equation}
\lambda_{mn,p} = t_{mn,p} + c_{mnp},
\end{equation}
where $t_{mn,p}$ satisfies the cyclic identity: $t_{[mn,p]} \equiv 0$ and $c_{mnp}$ is completely antisymmetric, we can write that $t_{mn,p} = \partial_m \omega_{np} - \partial_n \omega_{mp}$ with $\omega_{mn} = \omega_{nm}$(symmetric) and $c_{mnp} = \epsilon_{mnpq}\xi^q$ and then (\ref{A4}) lead to the two
gauge symmetries presented in \cite{DST}.

\end{document}